# The Security of Hardware-Based $\Omega(n^2)$ Cryptographic One-Way Functions: Beyond Satisfiability and P = NP

Javier A. Arroyo-Figueroa

Entevia, LLC [1]

email: jarroyo@entevia.com

## Abstract

We present a class of hardware-based cryptographic one-way functions that, in practice, would be hard to invert even if P=NP and linear-time satisfiability algorithms exist. Such functions use a hardware-based component with $\Omega(n^2)$ size circuits, and $\Omega(n^2)$ run time.

## 1. Introduction

Suppose someone has just proved that P=NP. A press conference is presented … the news travel at light speed through the world. Many emergency meetings are held at the major cryptographic security companies and government agencies. Malicious hackers start distributing linear-time decryption programs in the dark web. The end of the world is near! Well … not that fast.

In this paper, we prove that there is a class of cryptographic one-way functions that are, for all practical purposes, hard to invert even by the world's fastest computers, and even if P=NP: hardware-based $\Omega(n^2)$ cryptographic one-way functions.

The idea of hardware-based cryptography is not new. Plenty of research on the subject has been made during the past three decades. A good survey may be found in [G09].

Despite the intense research on hardware-based cryptography, such efforts have been concentrated in mainly two areas: (i) increase of performance by exploiting

---

hardware capabilities; and (ii) tamper-proof trapdoors, where the hardware provides tamper-proof methods for storing information essential for cryptographic functions that cannot be shared. Since the idea of using "slower" cryptographic functions seems counter-intuitive, we have not found any research effort that has been done in this aspect. This is the main contribution of our work.

The rest of this paper is organized as follows. In the next section, we present an overview of hardware-based $\Omega(n^2)$ cryptographic one-way functions. In Section 3, a typical application scheme is presented. The computational cost of using a linear-time SAT algorithm for finding the inverse of this class of functions is presented in Section 4. Finally, our conclusions are presented in Section 5.

## 2. Hardware-Based $\Omega(n^2)$ Cryptographic One-Way Functions

A hardware-based $\Omega(n^2)$ cryptographic one-way function (HCOWF2, for short) is a cryptographic one-way function that satisfies the following conditions:

1. It performs $\Omega(n^2)$ *unique* hardware-based computations; and
2. Each hardware-based computation is performed in $O(1)$ time by a $\Omega(n^2)$-size circuit.

The parameter n is typically called the "security parameter", or the "number of bits" of the function. A hardware-based computation is *unique* if the value of its inputs is not repeated in any subsequent computation, and the value of its outputs at each computation has $c/2^n$ probability of being repeated by any subsequent computation; where $c \geq 1$.

It is easy to implement a practical HCOWF2 using current hardware and software technology. For example, a 2048-bit function may take roughly $10^9$ instructions (assuming 100 instructions per computation), which would take 1 millisecond to compute with an Intel Celeron processor. In terms of hardware, such a function would roughly take 4 million gates, which can easily be implemented with any SRAM-based FPGA, and configured in less than a millisecond.

Contrary to what common sense may tell (e.g., look for linear lower bounds), we want an HCOWF2 to have a quadratic lower bound in both its hardware and software components. The motivation behind this is to make an HCOWF2 *practically-hard to invert* by current technology, even if P=NP. At the same time, we should be able to upgrade the security parameter as technology advances. It should be noted that this upgrade approach is not new, and is currently used in RSA encryption, where the security parameter is doubled in size every few years.

A quadratic lower bound in computational complexity can be achieved by making sure that a $\Omega(n)$ algorithm is executed n times, one for each of the n-bits specified by the security parameter. An example of this approach are Tau one-way functions [A16], which do n traversals through a bit matrix for every bit of its n-bit output.

A quadratic lower bound in the circuit complexity of the O(1) hardware-based computation can be achieved by constructing a *characteristic random function* H, whose logic circuit size is $\Omega(n^2)$ gates, specific to the HCOWF2. Typically, H returns an n-bit integer, and is invoked as H(p,q) where p is the input to the HCOWF2 (which could be a public key or a message digest), and q is a parameter from a set Q of random n-bit values, where $|Q| = n^2$.

The function H describes the $\Omega(n^2)$ hardware component. It is defined as H: $\{0,1\}^n$ x $\{0,1\}^n \rightarrow \{0,1\}^n$. The circuit for H can be generated by creating, for each output bit of H, a random 2n-variable Boolean formula from n irreducible k-CNF clauses, where k>2. Half of the 2n variables correspond to the "p" parameter, and the other half corresponds to "q". Let C be the set of all clauses that describe H, then each clause c in C is irreducible if there does not exist another clause c' in C with a set of variables common with c, that differ by only one literal. The total number of clauses is $\Omega(n^2)$, with a total number of gates in the circuit bounded by $\Omega(n^2)$.

## 3. Application Scheme

In a typical application scheme, both the sender and the receiver have a configurable, hardware-based H(p,q) component, as shown in Figure 1. While the sender stores its function description in a data store, the receiver maintains a cache of *function descriptions;* this avoids that senders be required to send their function description if it hasn't changed. Each function description includes the specification of the characteristic random function (H), as well the unique parameters to be passed to the function in each hardware-based computation. The description may also include other function-specific parameters as well.

On the sender's side, there is a one-time hardware configuration process, as shown for steps 1 and 2. The sender loads its function description from the data store, and configures the hardware component appropriately.

When sending a message, the sender runs its HCOWF2 by doing $\Omega(n^2)$ computations on the hardware component to build the message hash. Before sending the message, the sender sends a function-description signature to the receiver (step 4), which is used by the receiver to lookup the function description in its cache (step 5). In the case of a cache-miss, the receiver replies with a request for the sender to send its function description (step 6), to which the sender replies accordingly (step 7). In the case of a successful cache lookup in the receiver, steps 6 and 7 are not needed. The receiver then configures its hardware component using the function description.

Upon receiving a message and its authentication code (MAC) (step 9), the receiver verifies the MAC by doing $\Omega(n^2)$ computations on the hardware component, following the parameters of the function description.

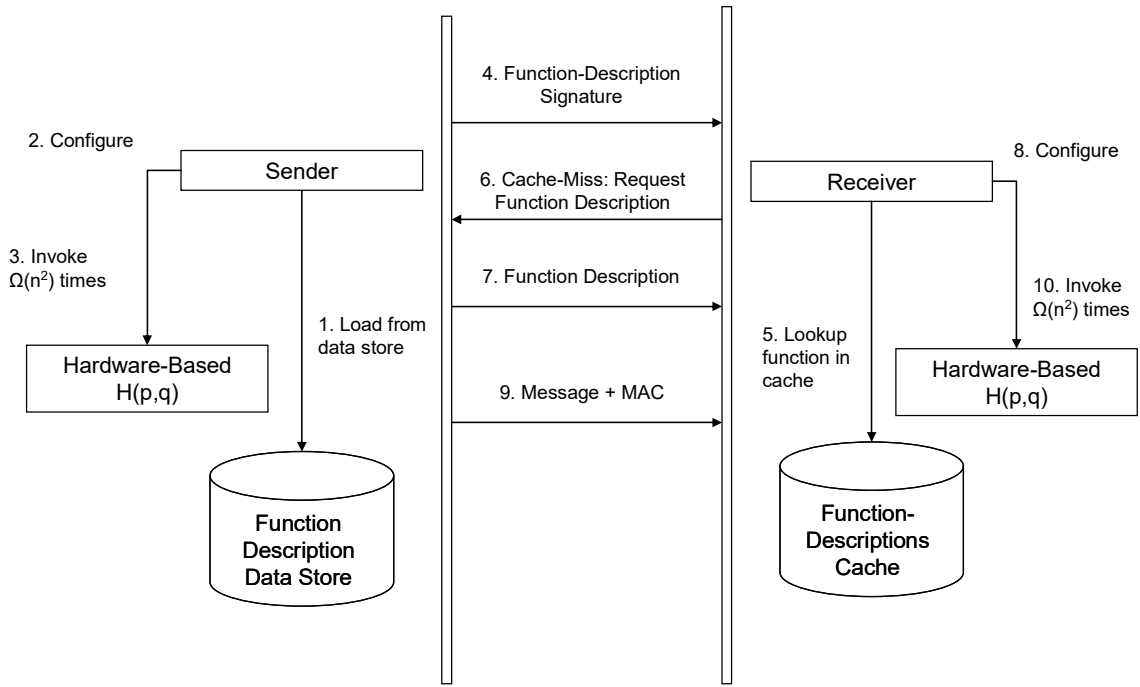

**Figure 1.** A typical application scheme

## 4. Computational Cost of Inversion with SAT if P=NP

Assuming P=NP, there exists a polynomial-time algorithm that solves the Boolean satisfiability (SAT) problem. It has been shown in [V04] that there is an almost-linear lower bound for a SAT algorithm, if P=NP. Thus, in order to invert a HCOWF2, it would be a matter of creating a Boolean circuit that represents the output, generate a k-CNF formula for the circuit, and run a linear-time SAT algorithm to determine the inputs to the formula. In this section, we will show that, even if this is the case, the computational cost of inverting HCOWF2 is high.

Given an n-bit output of a HCOWF2, a unique logic circuit represents each hardware-based computation (since each hardware-based computation is guaranteed to be unique). The number of gates of the circuit is bounded by $\Omega(n^2)$. Since there are $\Omega(n^2)$ unique hardware-based computations, a circuit that represents a HCOWF2 has a circuit whose total number of gates is bounded by $\Omega(n^4)$.

A Tseytin transformation [T70] provides the most compact 3-CNF representation of a logic circuit, with the drawback that there can be 3m total number of clauses, and 3m total number of extra variables, where m is the circuit size. Thus, there will be a total of $\Omega(3n^4+n)$ variables, and $\Omega(3n^4)$ clauses. Each literal will take $\Omega(\log(3n^4+n)+1)$ bits to be represented in each clause (the extra bit to indicate negation); and $\Omega(3\log(3n^4+n)+3)$ total bits per clause, as there are exactly three literals per clause. Therefore, the total number of bits needed to represent the Boolean formula for a HCOWF2 is bounded by $\Omega(3n^4 (3\log(3n^4+n)+3))$.

Let's do some math for our 2048-bit example. The size of a 3-CNF Boolean formula for a 2048-bit HCOWF2 will not be less than $3(2048)^4(3\log(3(2048)^4)+3) \approx 7.3 \times 10^{15}$ bits $\approx 9 \times 10^{14}$ bytes $\approx 828$ TB. That would be the total amount of memory needed to contain the Boolean formula for a single inversion. No single computing node has 828 TB of memory, thus a massively-parallel computer with thousands of nodes would be needed, where the formula is partitioned among the nodes to be able to fit the node's memory. Even for the case of P=NP, it has a great impact on the computational cost.

Since a massively-parallel computer is needed to solve SAT for the given formula, two approaches may be considered: formula partitioning and search-space partitioning. In the first case, the formula is partitioned optimally among the processing nodes, such that each node decides satisfiability on all the partitions; if all partitions are satisfiable, then the formula is satisfiable. In the case of search partitioning, each node evaluates the whole formula but in different search spaces, each one given by a subset of all possible variable assignments.

For the case of formula partitioning, if P=NP, a linear-time algorithm may exist to obtain an optimal partitioning. But such an algorithm must run a single node; otherwise, we will need to solve an optimal-partitioning-partitioning problem, to partition the partitioning problem into the computing nodes, and so forth. Thus a formula of size $3(2048)^4(3\log(3(2048)^4)+3)$ bits would need to be scanned at least once.

For our 2048-bit example, a minimum of approximately $9 \times 10^{14}$ bytes would need to be scanned in a single node. At current clock speeds of 4GHz, a minimum of $2.2 \times 10^5$ seconds $\approx 63$ hours would be needed.

For the case of search-space partitioning, the situation is not any better. In this approach, each processing node evaluates the formula against a subset of the variable assignments; then the formula is satisfiable if one node finds a satisfiable assignment within its search space. Since the whole formula needs to be scanned in each node, the minimum processing time is similar to formula partitioning.

For our 2048-bit example, the 63-hour minimum time is just for finding the satisfiability for a single-variable assignment. However, there are $\Omega(3n^4)$ variables in our formula, thus $\Omega(3n^4)$ executions of SAT decisions will be needed with self-reduction. Therefore, the minimum time required to find a satisfiable assignment is $64 \times 3(2048)^4 = 3.4 \times 10^{15}$ hours $= 3.8 \times 10^{11}$ years! And that's even for the case of P=NP and the assumption of the existence of linear-time algorithms for SAT and partitioning.

## 5. Conclusion

We have presented a class of hardware-based cryptographic one-way functions that are resilient to inversion by SAT algorithms, even if P=NP and such algorithms run in linear time in the world's fastest massively-parallel computers. The idea of such functions is to perform computations in a time bounded by $\Omega(n^2)$, and circuits bounded by $\Omega(n^2)$ in size. As a result, even with the application of compact transformations to Boolean formulas, such as Tseytin's, the resulting size of the formula makes it impractical to find the inverse of the function by any linear-time SAT algorithm.